\documentclass[titlepage,preprint,nofootinbib,prd,tightenlines,byrevtex,amsmath]{revtex4}
\usepackage[pctex32]{graphicx}
\usepackage{amsmath}

\newcommand{\beq}{\begin{equation}}
\newcommand{\eeq}{\end{equation}}
\newcommand{\eq}[1]{eq.(\ref{#1})}

\begin{document}
\preprint{UK-04-25}

\title {Three-Loop Radiative-Recoil Corrections to Hyperfine Splitting
in Muonium}
\author {Michael I. Eides}\email{eides@pa.uky.edu,
eides@thd.pnpi.spb.ru}
\affiliation{Department of Physics and Astronomy, University of
Kentucky, Lexington, KY 40506, USA,}
\affiliation{Petersburg Nuclear Physics
Institute, Gatchina, St.Petersburg 188350, Russia }
\author{Howard Grotch}\email{hgrotch@uky.edu}
\affiliation{Department of Physics and Astronomy, University of
Kentucky, Lexington, KY 40506, USA }
\author{Valery A. Shelyuto}\email{ shelyuto@vniim.ru}
\affiliation{D. I.  Mendeleev Institute of Metrology,
St.Petersburg 198005, Russia}


\begin{abstract}
We consider three-loop radiative-recoil corrections to hyperfine
splitting in muonium. These corrections are enhanced by the large
logarithm of the electron-muon mass ratio. The leading logarithm cubed
and logarithm squared contributions were obtained a long time ago. We
calculate the single-logarithmic and nonlogarithmic contributions of
order $\alpha^3(m/M)E_F$  generated by gauge invariant sets of
diagrams with one- and two-loop polarization insertions in diagrams
with two exchanged photons and radiative photons, and by  diagrams with
one-loop radiative photon insertions both in the electron and muon
lines. The results of this paper constitute a next step in the
implementation of the program of reduction of the theoretical
uncertainty of hyperfine splitting below 10 Hz. They  improve the
theory of hyperfine splitting, and affect the value of the
electron-muon mass ratio extracted from experimental data on the
muonium hyperfine splitting.
\end{abstract}



\maketitle

\section{Introduction: Leading Logarithmic Contributions of Order
$\alpha^2(Z\alpha)(\lowercase{m}/M)\widetilde E_F$}

\begin{figure}[ht]
\includegraphics[height=1.8cm]{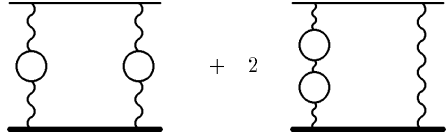}
\caption{\label{onelooppolrecfhsfig}
Graphs with two one-loop polarization insertions}
\end{figure}

The radiative-recoil corrections of order
$\alpha^2(Z\alpha)(m/M)\widetilde E_F$\footnote{We define the Fermi
energy  as
\beq      \label{baremuonfermi}
\widetilde{E}_{F}=\frac{16}{3}Z^4\alpha^2
\frac{m}{M} \left(\frac{m_r}{m}\right)^{3}ch\:R_{\infty},
\eeq
where $m$ and $M$ are the electron and muon masses, $\alpha$ is the
fine structure constant, $c$ is the velocity of light, $h$ is the
Planck constant, $R_{\infty}$ is the Rydberg constant, and $Z$ is the
nucleus charge in terms of the electron charge ($Z=1$ for muonium).
The Fermi energy  $\widetilde{E}_{F}$ does not include the muon
anomalous magnetic moment $a_\mu$ which does not factorize in the case
of recoil corrections, and should be considered on the same grounds as
other corrections to hyperfine splitting.} to hyperfine splitting in
muonium are enhanced by the large logarithm of the electron-muon mass
ratio cubed \cite{es0}. The leading logarithm cube contribution is
generated by the graphs in Fig.\ \ref{onelooppolrecfhsfig}
\footnote{And by the diagrams with the crossed exchanged photon
lines. Such diagrams with the crossed exchanged photon lines are
also often omitted in other figures below.} with insertions of the
electron one-loop polarization operators in the two-photon exchange
graphs. It may be obtained almost without any calculations by
substituting the effective charge $\alpha(M)$ in the leading recoil
correction of order $(Z\alpha)(m/M)\widetilde E_F$, and expanding the
resulting expression in the power series over $\alpha$ \cite{eks89}.

\begin{figure}[ht]
\includegraphics[height=1.8cm]{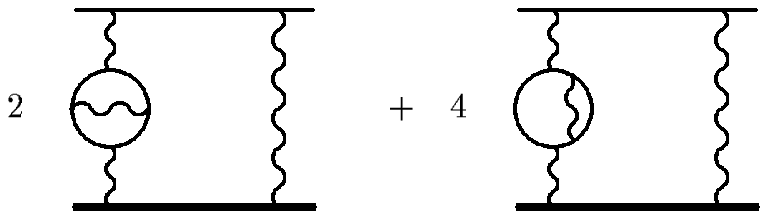}
\caption{\label{twolooppolrecfhsfig}
Graphs with two-loop polarization insertions}
\end{figure}

\begin{figure}[ht]
\includegraphics[height=2.2cm]{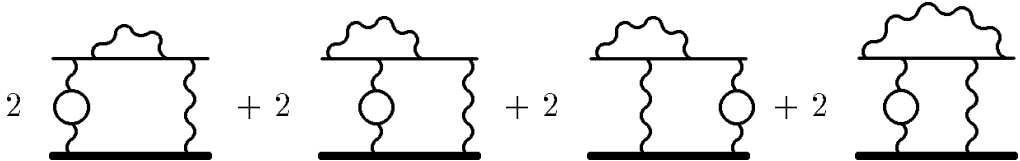}
\caption{\label{ellineinsrechfsfig}Graphs with radiative photon insertions}
\end{figure}

\begin{figure}[ht]
\includegraphics[height=2.1cm]{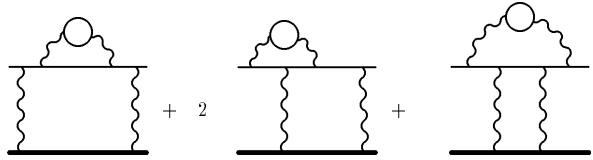}
\caption{\label{polellineinsrechfsfig} Graphs with polarization
insertions in the radiative photon}
\end{figure}

\begin{figure}[ht]
\includegraphics[height=2.5cm]{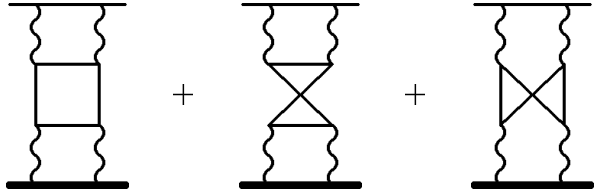}
\caption{\label{lihlightrechfsfig} Graphs with light by light
scattering insertions}
\end{figure}

\begin{figure}[ht]
\includegraphics[height=0.8cm]{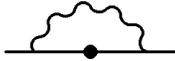}
\caption{\label{fifthcurhfsfig}Renormalization of the fifth current}
\end{figure}

Calculation of the logarithm squared term of order
$\alpha^2(Z\alpha)(m/M)\widetilde E_F$ is more challenging
\cite{eks89}. All graphs in Figs.\ \ref{onelooppolrecfhsfig},
\ref{twolooppolrecfhsfig}, \ref{ellineinsrechfsfig},
\ref{polellineinsrechfsfig}, and \ref{lihlightrechfsfig} generate
corrections of this order. The contribution induced by the irreducible
two-loop vacuum polarization in Fig.\ \ref{twolooppolrecfhsfig} is
again given by the effective charge expression.  Subleading logarithm
squared terms generated by the one-loop polarization insertions in
Fig.\ \ref{onelooppolrecfhsfig} may easily be calculated with the help
of the two leading asymptotic terms in the polarization operator
expansion and the skeleton integral. The logarithm squared contribution
generated by the diagrams in Fig.\ \ref{ellineinsrechfsfig} is obtained
from the leading single-logarithmic contribution of the diagrams without
polarization insertions by the effective charge substitution. An
interesting effect takes place in calculation of the logarithm squared
term generated by the polarization insertions in the radiative photon
in Fig.\ \ref{polellineinsrechfsfig}. One might expect that the high
energy asymptote of the electron factor with the polarization insertion
is given by the product of the leading constant term of the electron
factor $-5\alpha/(4\pi)$ and the leading polarization operator term.
However, this expectation turns out to be wrong.  One may check
explicitly that instead of the naive factor above one has to multiply the
polarization operator by the factor $-3\alpha/(4\pi)$. The reason for
this effect may easily be understood.  The factor $-3\alpha/(4\pi)$ is
the asymptote of the electron factor in massless QED and it gives a
contribution to the logarithmic asymptotics only after the polarization
operator insertion. This means that in massive QED the part
$-2\alpha/(4\pi)$ of the constant electron factor originates from the
integration region where the integration momentum is of order of the
electron mass. Naturally this integration region does not give any
contribution to the logarithmic asymptotics of the radiatively
corrected electron factor. The least trivial logarithm squared
contribution is generated by the three-loop diagrams in Fig.\
\ref{lihlightrechfsfig} with the insertions of the light by light
scattering block. Their contribution was calculated explicitly in
\cite{eks89}. Later it was realized that these contributions are
intimately connected with the well known anomalous renormalization of
the axial current in QED \cite{kes90}. Due to the projection on the HFS
spin structure in the logarithmic integration region, the heavy particle
propagator effectively shrinks to an axial current vertex, and in this
situation calculation of the respective contribution to HFS reduces to
substitution of the well known two-loop axial renormalization factor in
Fig.\ \ref{fifthcurhfsfig} \cite{adler} in the recoil skeleton diagram.
As expected, this calculation reproduces the same contribution as
obtained by direct calculation of the diagrams with light by light
scattering expressions. From the theoretical point of view it is
interesting that one can measure anomalous two-loop renormalization of
the axial current in the atomic physics experiment.

The sum of all logarithm cubed and logarithm squared  contributions of
order $\alpha^2(Z\alpha)(m/M)\widetilde E_F$ is given by the expression
\cite{es0,eks89}

\beq
\Delta
E=\left(-\frac{4}{3}\ln^3\frac{M}{m}+\frac{4}{3}\ln^2\frac{M}{m}\right)
\frac{\alpha^2(Z\alpha)}{\pi^3}\frac{m}{M} {\widetilde E}_F.
\eeq

\noindent
It was also shown in \cite{eks89} that there are no other
contributions with the large logarithm of the mass ratio squared
accompanied by the factor $\alpha^3$, even if the factor $Z$ enters in
another manner than in the equation above.

Single-logarithmic and nonlogarithmic terms of order
$\alpha^2(Z\alpha)(m/M)\widetilde E_F$ are generated by all diagrams in
Figs.\ \ref{onelooppolrecfhsfig}-\ref{polellineinsrechfsfig}, by the
graphs with the muon polarization loops, by the graphs with
polarization and radiative photon insertions in the muon
line, and also by the three-loop graphs with radiative photons in
the electron and/or muon lines.  Below we present our recent results
for three-loop single-logarithmic and nonlogarithmic  radiative-recoil
corrections.

\section{Two-Photon Exchange Diagrams. Cancellation of the
Electron and Muon Loops}

\begin{figure}[ht]
\includegraphics[height=1.5cm]{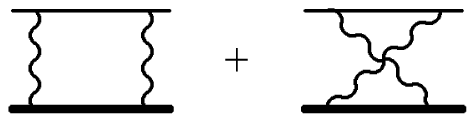}
\caption{\label{twophothfsfig}Diagrams with two-photon exchanges}
\end{figure}

Calculation of single-logarithmic and nonlogarithmic radiative-recoil
corrections of relative order $\alpha^2(Z\alpha)(m/M)$ (and also of
orders $(Z^2\alpha)^2(Z\alpha)(m/M)$ and
$\alpha(Z^2\alpha)(Z\alpha)(m/M)$) resembles in many respects
calculation of the corrections of relative orders
$\alpha(Z\alpha)(m/M)$ and $Z^2\alpha(Z\alpha)(m/M)$.
It was first discovered in \cite{ty,sty} that the contributions of the
diagrams with insertions of the electron and muon polarization loops
partially cancel, and, hence, it is convenient to treat such
diagrams simultaneously\footnote{We always consider the external muon
as a particle with charge $Ze$, this makes the origin of different
contributions more transparent. However, somewhat inconsequently we
omit the factor $Z$ in the case of the closed muon loops. The reason
for this  apparent inconsistency is just the cancellation which we
discuss now.}. Similar cancellation holds also for the corrections of
order $\alpha^2(Z\alpha)(m/M)\widetilde E_F$, so we will first remind
the reader how it arises when one calculates the polarization
contribution of order $\alpha(Z\alpha)(m/M)\widetilde E_F$. The
recoil contribution in the heavy particle pole of the two-photon
exchange diagrams exactly cancels in the sum of the electron and muon
polarizations (see for more details \cite{sty,egs01r}). Then the
skeleton recoil contribution to the hyperfine splitting generated by
the diagrams with two-photon exchanges in Fig.\ \ref{twophothfsfig} is
the result of the subtraction of the heavy pole contribution

\beq  \label{recskel}
\Delta E =4\frac{Z\alpha}{\pi}\frac{m}{M}
\widetilde{E}_F \int_{0}^{\infty}\frac{dk}{k}
\biggl[f(\mu k)-f\biggl(\frac{k}{2}\biggr)\biggr],
\eeq

\noindent
where $\mu=m/(2M)$, and

\beq
f(k)=\frac{1}{k}\biggl(\sqrt{1+k^2} - k -1\biggr)
- \frac{1}{2}\biggl( k\sqrt{1+k^2} - k^2- \frac{1}{2} \biggr),
\eeq
\[
f(k)_{k\to 0}\to -\frac{3}{4}+\frac{k^2}{2},\qquad f(k)_{k\to
\infty}\to -\frac{1}{k}.
\]

The electron polarization contribution is obtained from the skeleton
integral by multiplying the expression in \eq{recskel} by the
multiplicity factor 2, and inserting the polarization operator
$(\alpha/\pi)k^2I_1(k)$ in the integrand

\beq
\frac{\alpha}{\pi}k^2I_1(k)\equiv
\frac{\alpha}{\pi}k^2\int_0^1 {dv} \frac{v^2(1-v^2/3)}{4 + k^2(1-v^2)}.
\eeq

\begin{figure}[ht]
\includegraphics[height=1.5cm]{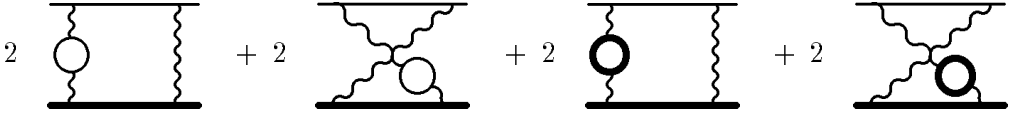}
\caption{\label{photlineradrechfsfir} Diagrams with one-loop
polarization insertions}
\end{figure}

\noindent
The muon polarization contribution is given by a similar expression,
and the total recoil contribution induced by the diagrams with both
the one-loop electron and muon polarizations in Fig.\
\ref{photlineradrechfsfir} has the form

\beq \label{electrmuoncontr}
\Delta E =8\frac{\alpha(Z\alpha)}{\pi^2}\frac{m}{M}
\widetilde{E}_F \int_{0}^{\infty}\frac{dk}{k}
\biggl[f(\mu
k)-f\biggl(\frac{k}{2}\biggr)\biggr][k^2I_1(k)+k^2I_{1\mu}(k)].
\eeq

\noindent
One can simplify this integral and demonstrate that with linear
accuracy in the small mass ratio $m/M$ all recoil contributions
generated by the diagrams with the one-loop electron and muon
polarization insertions in Fig.\ \ref{photlineradrechfsfir} are given
by the integral

\beq  \label{oneloopcanc}
\Delta E =8\frac{\alpha(Z\alpha)}{\pi^2}\frac{m}{M}
\widetilde{E}_F \int_{0}^{\infty}\frac{dk}{k}f(\mu k)k^2I_1(k).
\eeq

\noindent
This integral was calculated in \cite{sty} and we will not discuss its
calculation here. Our only goal in this Section was to demonstrate the
mechanism of the partial cancellation of the electron loop and muon
loop contributions.

\section{Diagrams with either Two Electron or Two Muon Loops}

\begin{figure}[ht]
\includegraphics[width=6cm]{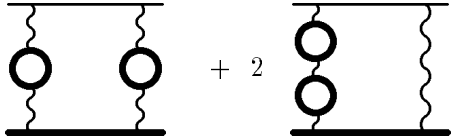}
\caption{\label{eloopmuloop} Graphs with two muon one-loop polarization
insertions}
\end{figure}

The nonrecoil contribution generated by the diagrams with two electron
or muon loops in Fig.\ \ref{onelooppolrecfhsfig} and Fig.\
\ref{eloopmuloop} was obtained a long time ago \cite{eks1}. Although it
was not emphasized in that work explicitly, it is easy to check that
the result in \cite{eks1} includes heavy pole contributions which are
due to the diagrams with both the electron and muon polarizations.
Repeating the same steps as in the previous Section, it is easy to see
that the recoil contribution generated by the diagrams in Fig.\
\ref{onelooppolrecfhsfig} and Fig.\ \ref{eloopmuloop} is determined by
the integral

\beq \label{twoelmuloopan}
\Delta E =12\frac{\alpha^2(Z\alpha)}{\pi^3}\frac{m}{M}
\widetilde{E}_F \int_{0}^{\infty}\frac{dk}{k}f(\mu k)k^4I_1^2(k),
\eeq

\noindent
where the numerical factor before the integral is due to the
multiplicity of the diagrams, and the whole integral is similar to the
integral in \eq{oneloopcanc}. The only significant difference is that
now we have the two-loop factor $k^4I^2(k)$ in the integrand instead of
the one-loop factor $k^2I_1(k)$. Calculating this integral we obtain
the total recoil contribution generated  by the diagrams in Fig.\
\ref{onelooppolrecfhsfig} and Fig.\ \ref{eloopmuloop} \cite{egs01}

\beq \label{finemuoneloop}
\Delta E
=\biggl[-\frac{4}{3}\ln^3{\frac{M}{m}}-\frac{8}{3}\ln^2{\frac{M}{m}}
- \biggl(\frac{2\pi^2}{3} + \frac{25}{9} \biggr)\ln{\frac{M}{m}}
- \frac{4\pi^2}{9} - \frac{535}{108}\biggr]
\frac{\alpha^2(Z\alpha)}{\pi^3}\frac{m}{M}\widetilde{E}_F.
\eeq

\noindent
The logarithm cube and logarithm squared terms in this expression were
obtained in \cite{es0,eks89}.

\section{Diagrams with both the Electron and Muon
Loops}\label{mixedidgrsect}

\begin{figure}[ht]
\includegraphics[width=6cm]{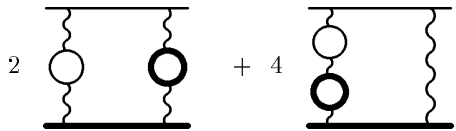}
\caption{\label{mixedloops}Graphs with both the electron and muon loops}
\end{figure}

Consider now the diagrams with one electron and one muon loop in Fig.\
\ref{mixedloops}. We can look at these diagrams as a result of the
electron polarization operator insertions in the muon loop diagrams in
Fig.\ \ref{photlineradrechfsfir}. The complete analytic expression for
the last two diagrams in Fig.\ \ref{photlineradrechfsfir} has the
form

\beq  \label{totalmuonone}
\Delta E =8\frac{\alpha(Z\alpha)}{\pi^2}\frac{m}{M}
\widetilde{E}_F \int_{0}^{\infty}\frac{dk}{k}
\biggl[\widetilde f(\mu k)-\widetilde f\biggl(\frac{k}{2}\biggr)
\biggr]k^2I_{1\mu}(k),
\eeq

\noindent
where

\[
\widetilde f(k)=f(k)+\frac{1}{k}.
\]

\noindent
After calculations we obtain  \cite{egs01}

\beq   \label{finmixoneloop}
\Delta E=
\biggl[
\biggl(\frac{2\pi^2}{3} - \frac{20}{9} \biggr) \ln{\frac{M}{m}}
+ \frac{\pi^2}{3} - \frac{53}{9} \biggr]\frac{\alpha^2(Z\alpha)}{\pi^3}
\frac{m}{M}\widetilde{E}_F.
\eeq

\section{Diagrams with Second Order Polarization
Insertions}\label{twolooppolsect}

\begin{figure}[ht]
\includegraphics[height=1.8cm]{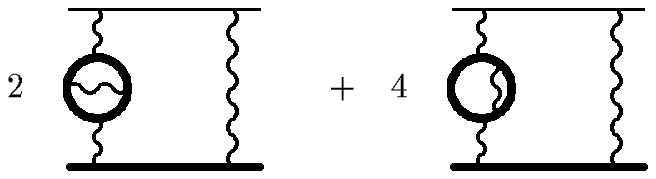}
\caption{\label{twoelmuloopanfig} Graphs with muon two-loop
polarization insertions}
\end{figure}

The recoil contribution to HFS generated by the diagrams in
Fig.\ \ref{twolooppolrecfhsfig} and Fig.\ \ref{twoelmuloopanfig} with
two-loop electron and muon polarization insertions is given by the
integral (compare \eq{oneloopcanc})

\beq  \label{twolooppolan}
\Delta E=
8\frac{\alpha^2(Z\alpha)}{\pi^3}\frac{m}{M}
\widetilde{E}_F \int_{0}^{\infty}\frac{dk}{k}f(\mu k)k^2I_2(k),
\eeq

\noindent
where $(\alpha^2/\pi^2)k^2I_2(k)$ is the two-loop polarization operator
\cite{kalsab,schwinger}

\beq
I_2(k) = \frac{2}{3} \int_0^1 {dv}\frac{v}{4+k^2(1-v^2)}
\biggl\{(3-v^2)(1+v^2)\biggl[\mbox{Li}_2 \biggl( -\frac{1-v}{1+v}\biggr)
\eeq
\[
+ 2\mbox{Li}_2 \biggl(\frac{1-v}{1+v}\biggr)
+\frac {3}{2} \ln{\frac{1+v}{1-v}} \ln{\frac{1+v}{2}}
- \ln{\frac{1+v}{1-v}} \ln{v} \biggr]
\]
\[
+ \biggl[\frac{11}{16}(3-v^2)(1+v^2) + \frac{v^4}{4}\biggr]
\ln{\frac{1+v}{1-v}}
\]
\[
+\biggl[\frac{3}{2}v(3-v^2)\ln{\frac{1-v^2}{4}}
- 2v(3-v^2)\ln{v} \biggr] + \frac {3}{8}v(5-3v^2)\biggl\}.
\]

The integral in \eq{twolooppolan} admits an analytic calculation and
the recoil contribution to HFS generated by the diagrams with
two-loop polarization insertions in Fig.\ \ref{twoelmuloopanfig} turns
out to be \cite{egs01}

\beq   \label{fintwoloop}
\Delta E= \biggl\{
-\frac{3}{2} \ln^2{\frac{M}{m}}
- \biggl[6\zeta (3) + \frac{13}{4}\biggr]\ln{\frac{M}{m}}
-\frac{97}{8}\zeta{(3)} - 16\mbox{Li}_4\biggl(\frac{1}{2}\biggr)
\eeq
\[
+ \frac{2\pi^2}{3}\ln^2{2} - \frac{2}{3}\ln^4{2}
+ \frac{5\pi^4}{36} - \frac{\pi^2}{4} + \frac{7}{16}\biggr\}
\frac{\alpha^2(Z\alpha)}{\pi^3}
\frac{m}{M}\widetilde{E}_F,
\]

\noindent
where $\zeta (3)=1.2020569\ldots$ and
$\mbox{Li}_4({1}/{2})=0.517479\ldots$. The logarithm squared
term in this expression was obtained in \cite{eks89}.

\section{Leading Four-Loop Radiative-Recoil Correction}

This leading four-loop radiative-recoil correction is generated by the
diagrams with four polarization operator insertions in the exchanged
photons similar to the diagrams with the three polarization
insertions in Fig.\ \ref{onelooppolrecfhsfig}. It contains the large
logarithm to the fourth power, and like the leading logarithm cubed
contribution in Eq.(2) may be easily obtained either by direct
calculation or by substitution of the effective charge $\alpha(M)$ in
the leading recoil correction of order $(Z\alpha)(m/M)\widetilde E_F$
\cite{egs01}

\beq
\Delta E=- \frac{8}{9}\ln^4{\frac{M}{m}}
\frac{\alpha^3(Z\alpha)}{\pi^4}~\frac{m}{M}{\widetilde E}_F.
\eeq

We consider this correction of order $\alpha^3(Z\alpha)$ together with
the radiative-recoil corrections of order $\alpha^2(Z\alpha)$ because,
due to the presence of the large logarithm, it is numerically of the
same order as these corrections

\beq  \label{fourlooplead}
\Delta E=-~1.~668 ~\frac{\alpha^2(Z\alpha)}{\pi^3}\frac{m}{M}{\widetilde
E}_F.
\eeq

\section{Diagrams with Radiative Photons in the Electron Line}

\subsection{Electron Vacuum Polarization}

Now we turn to consideration of diagrams which include both the
radiative photon insertions in the electron and/or muon line and the
polarization insertions in the exchanged photons. There are four gauge
invariant sets of such three-loop diagrams, and we calculate all their
contributions.

\begin{figure}[ht]
\includegraphics[height=1.9cm]{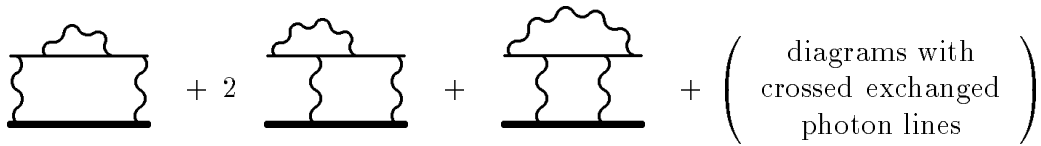}
\caption{\label{ellineradreclamb} Electron-line radiative-recoil
corrections}
\end{figure}

\begin{figure}[ht]
\includegraphics[height=1.7cm]{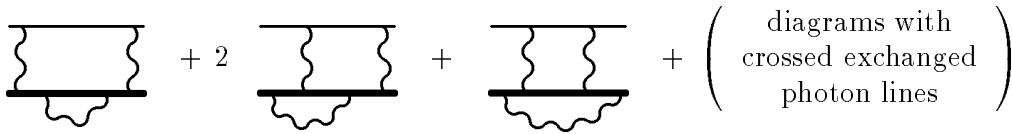}
\caption{\label{muonlineradrechfsfir} Muon-line radiative-recoil
corrections}
\end{figure}

All these four sets of diagrams can be obtained from
the two-photon exchange diagrams in Fig.\ \ref{ellineradreclamb} and
in Fig.\ \ref{muonlineradrechfsfir} with the radiative photons in the
electron or muon lines by insertions of the one-loop electron or muon
polarization operators.

\begin{figure}[ht]
\includegraphics[height=2cm]{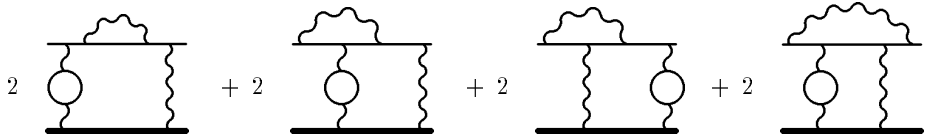}
\caption{\label{ee}Electron line and electron vacuum polarization}
\end{figure}

We have derived earlier a relatively compact integral representation
for the radiative-recoil corrections generated by the diagrams in
Fig.\ \ref{ellineradreclamb} and in Fig.\ \ref{muonlineradrechfsfir}
\cite{beks,egs98} (these integral representations do not include
anomalous magnetic moments of the electron and muon, respectively).
The explicit expression for this integral representation is too
cumbersome to reproduce it here.  The expression for the contribution
to HFS arising from the diagrams in Fig.\ \ref{ee} is obtained from the
integral in \cite{beks,egs98} by insertion in the integrand of the
doubled one-loop electron polarization $(\alpha/\pi)k^2I_e(k)$

\begin{equation}
2\left(\frac{\alpha}{\pi}\right)k^2I_e(k)=2\left(\frac{\alpha}{\pi}\right)
k^2\int_0^1dv\frac{v^2(1-v^2/3)}{4+k^2(1-v^2)},
\end{equation}

\noindent
where the additional multiplicity factor 2 corresponds to the fact
that we can insert the vacuum polarization in each of the exchanged
photons.

Calculating the respective integral we obtain the radiative-recoil
correction of order $\alpha^2(Z\alpha)(m/M)\widetilde E_F$ generated by
the diagrams in Fig.\ \ref{ee} (with the electron anomalous magnetic
moment subtracted) \cite{egs03}

\beq \label{eeresult}
\Delta E=
\left[\frac{5}{2} \ln^2{\frac{M}{m}}
+\frac{22}{3}\ln{\frac{M}{m}}+
11.41788~(3)\right]\frac{\alpha^2(Z\alpha)}{\pi^3}
\frac{m}{M}\widetilde{E}_F.
\eeq

\subsection{Muon Vacuum Polarization}

\begin{figure}[ht]
\includegraphics[height=2cm]{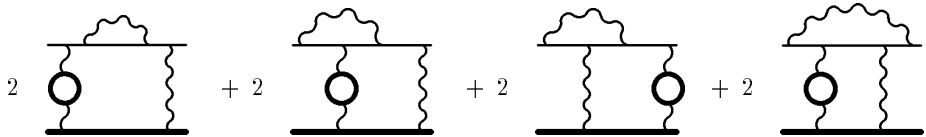}
\caption{\label{me} Electron line and muon vacuum polarization}
\end{figure}

Let us consider now the diagrams in Fig.\ \ref{me}. The only difference
between these diagrams and the diagrams in  Fig.\ \ref{ee} from  the
previous section is that they contain the muon vacuum polarization
insertion

\begin{equation}         \label{muoninsertioin}
2\left(\frac{\alpha}{\pi}\right)k^2I_\mu(k)
=2\left(\frac{\alpha}{\pi}\right)k^2
\int_0^1dv\frac{v^2(1-v^2/3)}{\mu^{-2}+k^2(1-v^2)},
\end{equation}

\noindent
instead of the electron vacuum polarization.

After calculations we obtain the contribution of the diagrams in Fig.\
\ref{me} (with the electron anomalous magnetic moment subtracted)
\cite{egs03}

\beq   \label{emuresult}
\Delta E=\left[-\frac{5\pi^2}{12} ~+~ \frac{1}{18}\right]
\frac{\alpha^2(Z\alpha)}{\pi^3}
\frac{m}{M}\widetilde{E}_F.
\eeq

\subsection{Electron Anomalous Magnetic Moment and Vacuum Polarizations}

It is well known that the electron anomalous magnetic moment
formally present in the diagrams in Fig.\ \ref{ellineradreclamb}
does not produce any contribution to the recoil corrections of order
$\alpha(Z\alpha)(m/M)\widetilde E_F$ \cite{stylett,sty,eksann1}. This
is the reason why the integral representation \cite{beks,egs98}
used above to calculate the corrections generated by the diagrams in
Fig.\ \ref{ee} and in Fig.\ \ref{me} does not include the anomalous
magnetic moment contributions. Such contributions connected with these
diagrams should be considered separately. As we discussed in Section II
there is a partial cancellation between the electron and muon loops in
the corrections with insertions of polarization loops in the exchanged
photons in Fig.\ \ref{photlineradrechfsfir}. We expect similar partial
cancellation in the electron anomalous magnetic moment contributions
generated by the diagrams in Fig.\ \ref{ee} and in Fig.\ \ref{me}, and
therefore consider the correction generated by the sum of these
diagrams. After calculations, we obtain an analytic result for the
electron anomalous magnetic moment contributions in  Fig.\ \ref{ee} and
in Fig.\ \ref{me}

\beq   \label{eammpol}
\Delta E=\left[-4\ln\frac{M}{m}+ \frac{4}{3}\right]
\frac{\alpha^2(Z\alpha)}{\pi^3}
\frac{m}{M}\widetilde{E}_F.
\eeq

\section{Diagrams with Radiative Photons in the Muon Line }

\subsection{Muon Vacuum Polarization}

\begin{figure}[ht]
\includegraphics[height=2cm]{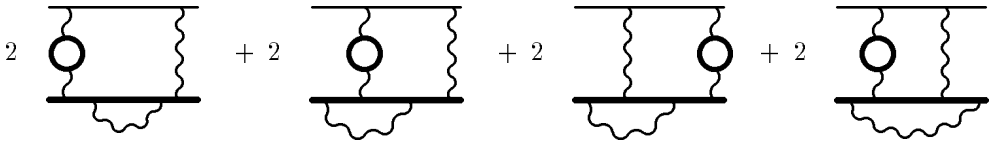}
\caption{\label{mm} Muon line and muon vacuum polarization}
\end{figure}

The contribution to the energy shift generated by the diagrams in Fig.\
\ref{mm} is obtained from the expression for the diagrams in Fig.\
\ref{muonlineradrechfsfir} by insertion in the integrand of the doubled
muon vacuum polarization

\begin{equation}         \label{muoninsertioinmu}
2\left(\frac{\alpha}{\pi}\right)k^2I_\mu(k)
=2\left(\frac{\alpha}{\pi}\right)k^2
\int_0^1dv\frac{v^2(1-v^2/3)}{4+k^2(1-v^2)}.
\end{equation}

\noindent
After calculations, we obtain the contribution to HFS generated by the
diagrams in Fig.\ \ref{mm} (with the muon anomalous magnetic
moment subtracted)  \cite{egs03}

\beq   \label{mumuresult}
\Delta E=-1.~80176~(2)\frac{\alpha(Z^2\alpha)(Z\alpha)}{\pi^3}
\frac{m}{M}\widetilde E_F.
\eeq

\subsection{Electron Vacuum Polarization}

\begin{figure}[ht]
\includegraphics[height=2cm]{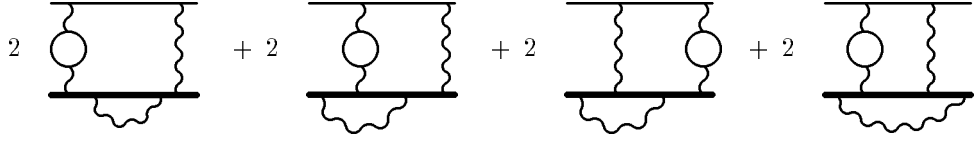}
\caption{\label{em} Muon line and electron vacuum polarization}
\end{figure}

Let us turn now to the diagrams in Fig.\ \ref{em}. The only difference
between these diagrams and the diagrams in Fig.\ \ref{mm} is that now we
have electron polarization insertions instead of the muon polarization
insertions.

The contribution to HFS generated by the diagrams in Fig.\
\ref{em} (with the muon anomalous magnetic  moment subtracted) is
equal to  \cite{egs03}

\beq  \label{mueresult}
\Delta  E=\left[\biggl(6\zeta{(3)} - 4 \pi^2 \ln{2} + \frac{13}{2}
\biggr) \ln{\frac{M}{m}} + 24.32115~(4)\right]
\frac{\alpha(Z^2\alpha)(Z\alpha)}{\pi^3} \frac{m}{M} \widetilde E_F.
\eeq

\subsection{Muon Anomalous Magnetic Moment and Vacuum Polarizations}

Consider now muon anomalous magnetic moment contributions
generated by the diagrams in Fig.\ \ref{mm} and in Fig.\ \ref{em}. As
in the case of radiative insertions in the electron line, we expect
partial cancellation (for more details see \cite{eksann2}) and
consider corrections generated by the sum of these diagrams. After
calculations, we obtain an analytic result for the muon anomalous
magnetic moment contributions in Fig.\ \ref{mm} and in Fig.\ \ref{em}

\beq   \label{muammpol}
\Delta E=\left[-4\ln\frac{M}{m}+ \frac{4}{3}\right]
\frac{\alpha(Z^2\alpha)(Z\alpha)}{\pi^3}
\frac{m}{M}\widetilde{E}_F.
\eeq

\section{Three-Loop Corrections Generated by One-Loop Fermion Factors}

\begin{figure}[ht]
\includegraphics[height=2cm]{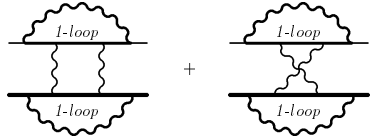}
\caption{\label{radrecdiag} Diagrams with two fermion factors}
\end{figure}

Consider three-loop radiative-recoil corrections to hyperfine splitting
in muonium generated by the diagrams in Fig.\ \ref{radrecdiag}.  These
corrections are nonlogarithmic and, unlike the case of the
radiative-recoil corrections of orders $\alpha(Z\alpha)(m/M)\widetilde
E_F$ and $(Z^2\alpha)(Z\alpha)(m/M)\widetilde E_F$, the one-loop
anomalous magnetic moments of both particles give nonvanishing
contributions to these diagrams. We use for calculations explicit
expressions for the subtracted electron and muon factors obtained
earlier \cite{beks,egs98}.

After tedious calculations, we obtain an analytic result for the
three-loop radiative-recoil correction to hyperfine splitting in
muonium generated by the diagrams in Fig.\ \ref{radrecdiag} \cite{egs04}

\beq \label{newrsult}
\Delta E =\biggl[-\frac{15}{8}\zeta{(3)} + \frac{15\pi^2}{4}
\ln{2} + \frac{27\pi^2}{16} - \frac{147}{32} \biggr]
\frac{\alpha (Z^2\alpha)(Z\alpha)}{\pi^3}\frac{m}{M}\,
~\widetilde E_F.
\eeq

\section{Discussion of Results}

Combining all three-loop single-logarithmic and nonlogarithmic
corrections to hyperfine splitting in \eq{finmixoneloop},
\eq{fintwoloop}, \eq{fourlooplead}, \eq{eeresult}, \eq{emuresult},
\eq{eammpol} \eq{mumuresult}, \eq{mueresult}, \eq{muammpol}, and
\eq{newrsult} we obtain  ($Z=1$ below)

\beq
\Delta E_{tot}=\biggl[\biggl(- 4 \pi^2\ln{2}
- \frac{29}{12}\biggr) \ln{\frac{M}{m}}
- 14\zeta(3)- 16\mbox{Li}_4\biggl(\frac{1}{2}\biggr)
+\frac{2\pi^2}{3}\ln^22
\eeq
\[
+ \frac{15\pi^2}{4} \ln{2}-\frac{2}{3} \ln^42+ \frac{5\pi^4}{36}
+ \frac{191\pi^2}{144}
-\frac{10655}{864}+29.88049~(6)\biggr]\frac{\alpha^3}{\pi^3}\frac{m}{M}
~\widetilde E_F,
\]

\noindent
or

\begin{equation}
\Delta E_{tot}=\biggl[\biggl(- 4 \pi^2\ln{2}
- \frac{29}{12}\biggr) \ln{\frac{M}{m}}
+47.7213\biggr]
\frac{\alpha^3}{\pi^3}\frac{m}{M}
~\widetilde E_F.
\end{equation}

\noindent
Numerically this contribution to the muonium HFS is

\begin{equation}
\Delta E_{tot}=-0.030~4~\mbox{kHz}.
\end{equation}

Currently the theoretical accuracy of hyperfine splitting
in muonium is about 70 Hz. A realistic goal is to reduce this
uncertainty below 10 Hz (see a more detailed discussion in
\cite{egs01r,egs03}).  The contributions discussed above,
\eq{newrsult}, together with the results of other recent research
\cite{egs98,my,rh} makes achievement of this goal closer.
Phenomenologically, the improved accuracy of the theory of hyperfine
splitting would  lead to a reduction of the uncertainty of the value of
the electron-muon mass ratio derived from the experimental data
\cite{lbdd} on hyperfine splitting (see, e.g., reviews in
\cite{egs01r,mt00}).

The single-logarithmic and nonlogarithmic three-loop radiative-recoil
corrections generated by the gauge invariant sets of diagrams with
two-loop fermion factors and light-by-light insertions in the
exchanged photons remain to be calculated. Work on their calculation
is in progress now.

\vskip0.5cm

{\bf Acknowledgments}

\vskip0.5cm

This work was supported in part by the NSF grant PHY-0138210.  The work
of V.  A. Shelyuto was also supported in part by the RFBR grants
03-02-04029 and 03-02-16843 and DFG grant GZ 436 RUS 113/769/0-1.


\begin{thebibliography}{99}

\bibitem{es0} M. I. Eides and V. A. Shelyuto, Phys. Lett. {\bf{146B}},
241 (1984).

\bibitem{eks89} M. I. Eides, S. G. Karshenboim, and V. A. Shelyuto,
Phys. Lett. {\bf{216B}}, 405 (1989); Yad. Fiz. {\bf{49}}, 493 (1989) [Sov.
J. Nucl. Phys. {\bf{49}}, 309 (1989)].

\bibitem{kes90} S. G. Karshenboim, M. I. Eides, and V. A. Shelyuto, Yad.
Fiz.  {\bf{52}}, 1066 (1990) [Sov. J. Nucl.  Phys. {\bf{52}} (1990)
679].

\bibitem{adler} S. L. Adler, Phys. Rev. {\bf 177}, 2426 (1969).

\bibitem{ty} E. A. Terray and  D. R. Yennie, Phys. Rev. Lett.
{\bf 48}, 1803 (1982).

\bibitem{sty} J. R. Sapirstein, E. A. Terray, and D. R. Yennie, Phys. Rev.
{\bf D29}, 2290 (1984).

\bibitem{egs01r} M. I. Eides, H. Grotch, and V. A. Shelyuto, Phys. Rep.
{\bf 342}, 63 (2001).

\bibitem{eks1}  M. I. Eides, S. G. Karshenboim, and V. A. Shelyuto, Phys.
Lett. {\bf{229B}},  285 (1989); Pis'ma Zh. Eksp. Teor. Fiz. {\bf{50}}, 3
(1989) [JETP Lett.  {\bf{50}}, 1 (1989)]; Yad. Fiz. {\bf{50}},
1636 (1989) [Sov. J. Nucl. Phys.{\bf{50}}, 1015 (1989)].

\bibitem{egs01} M. I. Eides, H. Grotch, and V. A. Shelyuto, Phys. Rev.
D {\bf 65}, 013003 (2002).

\bibitem{kalsab}  G. Kallen and A. Sabry, Kgl.  Dan.  Vidensk.  Selsk.
Mat.-Fis.  Medd.  {\bf 29} (1955) No.17.

\bibitem{schwinger} J.  Schwinger,  Particles, Sources and Fields,  Vol.2
(Addison-Wesley, Reading, MA, 1973).

\bibitem{beks}  V. Yu. Brook, M. I. Eides, S. G. Karshenboim, and
V. A. Shelyuto, Phys. Lett. B {\bf{216}}, 401 (1989).

\bibitem{egs98}  M. I. Eides, H. Grotch, and V. A. Shelyuto, Phys. Rev.
D {\bf 58}, 013008 (1998).

\bibitem{egs03} M. I. Eides, H. Grotch, and V. A. Shelyuto, Phys. Rev.
D {\bf 67}, 113003 (2003).

\bibitem{stylett} J. R. Sapirstein, E. A. Terray, and D. R. Yennie, Phys.
Rev. Lett. {\bf 51}, 982 (1983).

\bibitem{eksann1} M. I. Eides, S. G. Karshenboim, and V. A. Shelyuto, Ann.
Phys.  (NY) {\bf 205}, 231 (1991).

\bibitem{eksann2} M. I. Eides, S. G. Karshenboim, and V. A. Shelyuto,
Ann.  Phys.  (NY) {\bf 205}, 291 (1991).

\bibitem{egs04} M. I. Eides, H. Grotch, and V. A. Shelyuto,  Phys. Rev.
D {\bf 70}, 073005 (2004).

\bibitem{my} K. Melnikov and A. Yelkhovsky, Phys. Rev. Lett. {\bf 86},
1498 (2001).

\bibitem{rh} R. J. Hill, Phys. Rev. Lett. {\bf 86}, 3280 (2001).

\bibitem{lbdd} W. Liu, M. G. Boshier, S. Dhawan et al, Phys. Rev. Lett. {\bf
82}, 711 (1999).

\bibitem{mt00} P. J. Mohr and B. N. Taylor, Rev. Mod. Phys. {\bf 72},
351 (2000).


\end{thebibliography}
\end{document}